\documentclass[a4paper,10pt]{article}
\usepackage[english]{babel}
\usepackage[T1]{fontenc}
\usepackage[utf8]{inputenc}
\usepackage{amssymb}
\usepackage{physics}
\usepackage{amsmath}
\usepackage{graphicx}
\usepackage[unicode]{hyperref}
\usepackage{tikz}
\usepackage{color}
\usepackage{genyoungtabtikz}
\usepackage[title]{appendix}
\usepackage{indentfirst}
\usepackage{bm}
\usepackage{xhfill}
\usepackage{bbm}
\usepackage{multirow}
\usepackage{array}
\usepackage[centertableaux]{ytableau}
\usepackage{multicol}
\usepackage[shortlabels]{enumitem}
\usepackage{forest}
\usepackage{float}

\hypersetup{
    colorlinks=true,
    linkcolor=blue,
    filecolor=magenta,      
    urlcolor=cyan,
}
 
\usepackage{amsthm}

\theoremstyle{definition}

\usepackage[nottoc]{tocbibind}

\usepackage{authblk}

\title{\textbf{Fragmented perspective of self-organized criticality and disorder in log gravity}}

\author[1,2]{\textbf{Yannick Mvondo-She}}
\affil[1]{School of Physics and Mandelstam Institute for Theoretical Physics,
University of the Witwatersrand, Johannesburg, Wits, 2050, South Africa}
\affil[2]{National Institute of Theoretical and Computational Sciences, Private Bag X1, Matieland, South Africa}
\affil[ ]{\texttt{yannick.mvondo-she@nithecs.ac.za}}

\date{}

\begin{document}

\maketitle

\begin{abstract}
We use a statistical model to discuss nonequilibrium fragmentation phenomena taking place in the stochastic dynamics of the log sector in log gravity. From the canonical Gibbs model, a combinatorial analysis reveals an important aspect of the $n$-particle evolution previously shown to generate a collection of random partitions according to the Ewens distribution realized in a disconnected double Hurwitz number in genus zero. By treating each possible partition as a member of an ensemble of fragmentations, and ensemble averaging over all partitions with the Hurwitz number as a special case of the Gibbs distribution, a resulting distribution of cluster sizes appears to fall as a power of the size of the cluster. Dynamical systems that exhibit a distribution of sizes giving rise to a scale-invariant power-law behavior at a critical point possess an important property called self-organized criticality. As a corollary, the log sector of log gravity is a self-organized critical system at the critical point $\mu l =1$. A similarity between self-organized critical systems, spin glass models and the dynamics of the log sector which exhibits aging behavior reminiscent of glassy systems is pointed out by means of the P\`{o}lya distribution, also known to classify various models of (randomly fragmented) disordered systems, and by presenting the cluster distribution in the log sector of log gravity as a distinguished member of this probability distribution. We bring arguments from a probabilistic perspective to discuss the disorder in log gravity, largely anticipated through the conjectured AdS$_3$/LCFT$_2$ correspondence.
\end{abstract}

\tableofcontents

\section{Introduction}
Fragmentation is a fundamental stochastic process that produces a cluster distribution of objects, and is observed in areas of science as disparate as relativistic heavy-ion collisions \cite{chase1994nuclear}, spin glass models \cite{mezard1987spin}, or population genetics \cite{mekjian1991cluster,higgs1995frequency,derrida1999genealogical}. In this paper, We present a statistical study of the fragmentation and clustering processes that occur in the logarithmic sector of cosmological topologically massive gravity at the critical point, and give a perspective of the fragmentation-associated disorder in the theory. 

Cosmological topologically massive gravity at the critical point (CCTMG), is a special model of three-dimensional gravity theory in anti-de Sitter (AdS) space-time that saw the light of day in 2008 \cite{Grumiller:2008qz}. Also called log gravity \cite{Maloney:2009ck}, this theory is equipped with a negative cosmological constant as well as a gravitational Chern-Simons term, and has the particularity that under a more relaxed class of boundary conditions than the standard Brown-Henneaux ones \cite{Brown:1986nw}, it features a new physical degree of freedom that behaves as a complex logarithmic function, the so-called logarithmic primary mode  

 \begin{eqnarray}
 \label{log mode}
 \psi^{new}_{\mu \nu} := \lim_{\mu l \rightarrow 1}  \frac{\psi^M_{\mu \nu} (\mu l) - \psi^L_{\mu \nu}}{\mu l -1} = \left\{ -\ln \cosh{\rho} - i \tau \right\} \psi^L_{\mu \nu},
 \end{eqnarray}

\noindent which under the action of the $\rm{SL \left( 2, \mathbb{R} \right) \times SL \left( 2, \mathbb{R} \right)}$ AdS$_3$ isometry group generates descendant single particle logarithmic modes, and whose logarithmic branches extend the single particle class of solutions to a logarithmic multi-particle sector. The logarithmic primary mode spoils the unitarity of the theory, but rather than a bug, this aspect of the theory appeared to be of great interest as it revealed the salient features of a potential holographic counterpart, logarithmic conformal field theory \cite{Grumiller:2013at}, well-known for describing disordered systems.

A study of the partition function was initiated in \cite{Gaberdiel:2010xv} and systematically extended in a series of papers \cite{Mvondo-She:2018htn,Mvondo-She:2019vbx,Mvondo-She:2021joh,Mvondo-She:2022jnf,Mvondo-She:2023xel,Mvondo-She:2023ppz}. From \cite{Gaberdiel:2010xv}, the calculation of the 1-loop graviton partition function of log gravity on the thermal AdS$_3$ background resulted in the expression

\begin{eqnarray}
\label{z tmg}
{Z_{\textrm{CCTMG}}} (q, \bar{q})= \prod_{n=2}^{\infty} \frac{1}{|1-q^n|^2} \prod_{m=2}^{\infty} \prod_{\bar{m}=0}^{\infty} \frac{1}{1-q^m \bar{q}^{\bar{m}}}, \hspace{1cm} \text{with} \hspace{0.25cm} q=e^{2i \pi \tau}, \bar{q}=e^{-2i \pi \bar{\tau}},
\end{eqnarray}

\noindent where the first product can be identified as the three-dimensional gravity partition function $Z_{0,1}$ in \cite{Maloney:2007ud}, while the second product contributes to the added logarithmic sector. The calculation of the one-loop partition function showed agreement with the partition function of a logarithmic conformal field theory up to single particle, while an interpretation of the multiparticle sector remained elusive. By writing Eq. (\ref{z tmg}) as 

\begin{eqnarray}
\label{z grav z log}
Z_{\textrm{CCTMG}} (q, \bar{q})=  Z_{gravity} (q, \bar{q}) \cdot Z_{log} (q, \bar{q}),
\end{eqnarray}
where

\begin{eqnarray}
Z_{gravity} (q, \bar{q})= \prod_{n=2}^{\infty} \frac{1}{|1-q^n|^2}, \hspace{0.5cm} \mbox{and}
\hspace{0.5cm} Z_{log} (q, \bar{q}) = \prod_{m=2}^{\infty} \prod_{\bar{m}=0}^{\infty} \frac{1}{1-q^m \bar{q}^{\bar{m}}},
\end{eqnarray}

\noindent subsequent studies \cite{Mvondo-She:2018htn,Mvondo-She:2019vbx,Mvondo-She:2021joh,Mvondo-She:2022jnf,Mvondo-She:2023xel,Mvondo-She:2023ppz} endeavored to systematically study the partition function of the log sector $Z_{log} (q, \bar{q})$, in order to find a unifying interpretation for both single and multiparticle sectors. A summary of the additional features of the theory that are of interest for this paper is listed below.

\begin{itemize}
\item The log partner is a complex dynamic field which together with its descendant fields form collective fields whose $n$-particle evolution is encoded in the log partition function by collecting the fields under a rescaling of variables with the coordinate sequence $\left( \mathcal{G}_k \right)_{k=1}^n$ as 

\begin{eqnarray}
\label{G_n}
\mathcal{G}_k \left( q,\bar{q} \right) = \frac{1}{|1-q^k|^2}, 
\end{eqnarray}

\noindent such that 

\begin{eqnarray}
Z_{log} \left( \mathcal{G}_1, \ldots , \mathcal{G}_n  \right) &=& 1 + \sum_{n=1}^{\infty}  \left( \sum_{\substack{\sum_{k=1}^n k j_k =n \\ n \geq 1\\j_k \geq 0 }} \left\{ H^{\bullet}_{0 \xrightarrow[]{n} 0} \left[ \left( [k]^{j_k} \right)_{k=1}^n,  \left( [k]^{j_k} \right)_{k=1}^n \right] \right\} \prod_{k=1}^n \mathcal{G}_k^{j_k} \right) \left( q^2 \right)^n, \nonumber \\ && \label{Hurwitz tau function}   
\end{eqnarray}

\noindent and where the expression

\begin{eqnarray}
\label{Hurwitz numbers}
H^{\bullet}_{0 \xrightarrow[]{n} 0} \left[ \left( [k]^{j_k} \right)_{k=1}^n,  \left( [k]^{j_k} \right)_{k=1}^n \right] = \prod_{k=1}^n \frac{1}{j_k!  (k)^{j_k}},
\end{eqnarray}

\noindent given in term of the sequence  $\left( [k]^{j_k} \right)_{k=1}^n = \left( [1]^{j_1}, \ldots, [n]^{j_n} \right)$ associated to the monomials $\prod_{k=1}^n \mathcal{G}_k^{j_k}$ with $[k]^{j_k}= \overbrace{k, \ldots. k}^{j_k ~ \rm{times}}$, is a disconnected double Hurwitz number in genus zero counting $n$-fold coverings of $\mathbb{CP}^1$ by itself.

\item The instability originally found in \cite{Grumiller:2008qz} was presented as of a solitonic type by viewing the coordinate sequence $\left( \mathcal{G}_k \right)_{k=1}^n$ as an ensemble of flows or time variables, and showing that $Z_{log} \left( \mathcal{G}_1, \ldots , \mathcal{G}_n  \right)$ is a $\tau$-function of the KP I hierarchy.

\item The description of the log sector as a KP dynamical integrable system was shown to be provided with a physically consistent measure, the Ewens sampling formula \cite{ewens1972sampling}, revealing a stochastic process in the $n$-particle evolution.

\item The stochastization of the dynamical process in terms of the Ewens sampling formula allowed to model the evolution of the $n$-particle log sector in terms of the Hoppe urn model \cite{hoppe1984polya}, which shows that the configuration of colored balls generated by a black ball after $n$ draws from the urn is distributed as the Ewens sampling formula.

\item The urn model provides a stochastic mechanism for generating the class of random permutations of the Ewens distribution that is in one-to-one correspondence with genus-zero Feynman diagrams, i.e rooted trees, with the particular correspondence that the balls in the urn model are vertices in trees and the black ball which as a device for introducing new mutations is ignored in describing the urn configuration, corresponds to the root of trees.

\item The $n$-particle evolution generates a particular type of Markov process, a random walk on the symmetric group manifold. Through an equivalence between any gauge theory on a two-dimensional surface of genus zero and a random walk on the gauge group manifold, the log sector of the theory, via the Hurwitz number in its log partition which describes the statistics of branched $n$-coverings of Riemann surfaces, was shown to be a two-dimensional gauge theory of the symmetric group.
\end{itemize}

From the above, it is our belief that besides randomness, a non-equilibrium process takes place in the theory, and the goal of this paper is to make this point clear. 

In section 2, we show that the fragmentation process in the $n$-particle's log partition evolution can be described by a statistical Gibbs model and belongs to a class of exactly solvable models extensively used to analyse the fragmentation of nuclei in heavy-ion collisions \cite{mekjian1990model,mekjian1990distribution,mekjian1991models}. The $n$-particle fragmentation that occurs in the log sector is shown to be a special case of the statistical model, which involves permutations and their cycle class decomposition. A permutation is specified by listing the number of cycles with associated cycle lengths that make up the permutation. In this case, the number $j_k$ of clusters of size $k$ is just the number of cycles of length $k$. We then give a proof that the statistic number of $k$-cycles takes an average value $\langle j_k \rangle$ that falls as a $1/k$ scale invariant power-law. Such scaling behavior related to size distribution in an evolving unstable spectrum at a particular critical point has widely been discussed in the literature, particularly in the context of sandpile slides and models of self organized criticality \cite{bak1987self,bak1988self} or avalanche slides \cite{kadanoff1989scaling}. The appearance of a power law behavior by ensemble averaging $j_k$ is a remarkable feature of the distribution of clusters which signals self-organized criticality in log gravity. Self-organised criticality (SOC) was introduced in \cite{bak1987self}, in order to address out-of-equilibrium driven dynamical systems displaying power law behavior, fractal geometry, and scale invariance characteristic of the critical point of phase transition, and has found applications in many areas of modern scientific research such as in geophysics, neuro- and evolutionary biology, cosmology, astrophysics, collective human (traffic flow, stock markets) and animal behavior, cellular automation. The paradigmatic illustration of a self-organized critical system is the sandpile model, which can be described as a model with a stochastic dynamics in which a sand grain is dropped randomly on a sandpile, subsequently provoking a downward avalanche of grains, when the system is driven to some critical condition.

In section 3, a similarity between spin glass models, a well-known instance of SOC system \cite{le2010avalanches} and (the log sector of) log gravity is pointed out. The parallel is motivated by the fact that the log sector is modeled as a hierarchical tree exhibiting an aging behavior reminiscent of glassy disordered systems, in similar fashion as SOC systems with glassy dynamics in which a (random) hierarchical tree structure emerge dynamically \cite{bak1997self}. A probabilistic classification of the disorder in log gravity is presented by means of the general P\'{o}lya distribution, also known for categorizing various disordered systems, among spin glass models. The cluster distribution in the log sector of log gravity is clearly shown to be a special case of the P\'{o}lya distribution. By showing how the log partition function can be expressed in terms of a limit of the P\'{o}lya distribution, as well as by showing how the connected Hurwitz number plays the role of the distribution, we bring arguments to discuss the disorder in log gravity, largely anticipated through the conjectured AdS$_3$/LCFT$_2$ correspondence.

In section 4, we summarize our results and propose a correspondence between log gravity and the sandpile model. 

\section{Fragmentation, power-law behavior and self-organized criticality}
In order to discuss the fragmentation of the log $n$-particle sector into clusters, we consider a partitioning of $n$ into groups of varying sizes which gives rise to a distribution of cluster sizes.

Let $j_i$ be the number of clusters of size $i$. We the partition 

\begin{eqnarray}
\Vec{j}= \left( j_1,j_2, \ldots , j_n \right),    
\end{eqnarray}

\noindent constrained by 

\begin{eqnarray}
\label{constraint}
n = \sum_{k=1}^{n} k j_k,    
\end{eqnarray}

\noindent and the multiplicity $m$ as the number of clusters in a particular partition with expression 

\begin{eqnarray}
m = \sum_{k=1}^{n} j_k.   
\end{eqnarray}

\noindent Such partitioning also appears in the classification of permutations by cycle classes where a cycle class is specified by $\left( j_1,j_2, \ldots, j_i,\ldots \right)$ and $j$ specifies the length of a cycle in the permutation case.

\subsection{Statistical fragmentation model}

A statistical method relating the counting of a system of identical particles which forms clusters of various sizes to the counting of a set of states represented by integer partitions can be formulated by assigning a weight to each partition $\Vec{j}$ and obtaining the distribution of clusters by ensemble averaging $n$, over all partitions using the
weight.

Consider the $j_k$ identical particles of type $k$ to occupy a group of $x_k$ states, and suppose the probability of forming a $k$-cluster to be proportional to $x_k$. If these probabilities were independent (which is not the case), the weight function $W \left(  \Vec{j} \right)$ giving the number of states of $j_k$ identical particles of type $k$ occupying the group of $x_k$ states would have the proportionality relationship
$W \left(  \Vec{j} \right) \propto \prod_k x_k^{j_k}$. However, in forming the $j_k$ $k$-clusters, there is an overcounting problem associated to the indistinguishability of the clusters, which requires the weight to be adjusted according to a rule also known as the "correct Boltzmann counting" \cite{huang2008statistical}. The modification consists in reducing the size of the phase space available to the fragments by a factor of $1/j_k!$. The correct weight therefore reads

\begin{eqnarray}
\mathcal{W} \left(  \Vec{j}, \Vec{x} \right) = \prod_{k > 0}  \frac{x_k^{j_k}}{j_k!},  
\end{eqnarray}

\noindent where $\vec{x}$ is a vector of parameters, and defines a canonical Gibbs model.

The fundamental law of statistical clustering theories states that out of the competition between fragmentation as the process through which an object breaks up into smaller pieces and clustering as the inverse process of objects combining to form larger parts, a distribution of sizes emerges.

A probability distribution of cluster sizes associated with the Gibbs model can be arrived at by specifying the Gibbs weight with the particular choice $x_k= x/k$, with $x$ as a single parameter describing the degree of fragmentation, and writing the Gibbs distribution as

\begin{eqnarray}
\label{Gibbs distribution}
\mathcal{P}_n \left( \Vec{j}, x   \right)= \frac{n!}{Q_n \left( x \right)} \mathcal{W} \left(  \Vec{j}, x \right) = \frac{n!}{Q_n \left( x \right)} \prod_{k=1}^n \frac{1}{j_k!} \left(  \frac{x}{k} \right)^{j_k},    
\end{eqnarray}

\noindent satisfying the probability condition

\begin{eqnarray}
\sum_{k=1}^{n} \mathcal{P}_n \left( \Vec{j}, x   \right)= 1,    
\end{eqnarray}

\noindent with 

\begin{eqnarray}
Q_n(x) = x (x+1) (x+2) \cdots (x+n-1) = \frac{\Gamma (x+n)}{\Gamma (x)}.    
\end{eqnarray}

\noindent Such a probability distribution became an extensively used analytical tool in the modeling of nuclear fragmentation  \cite{mekjian1990model,mekjian1990distribution,mekjian1991models,lee1992mass,lee1992canonical}. It also appears in population genetics as the Ewens sampling formula \cite{ewens1972sampling} for the distribution of alleles \cite{mekjian1991cluster,chase1994nuclear}. 

We also note that at a more formal level, the family of Ewens distributions on set partitions is a distinguished one among the class of so-called Canonical Gibbs distribution \cite{crane2016ubiquitous}.

\subsection{Power-law behavior and self-organized criticality}
Self-organized criticality \cite{bak1987self}, is a phenomenon that features an out-of-equilibrium driven dynamic system evolving to reach a stationary state through a spatial and temporal multi-scale response characterized by a cascade of events (or an avalanche). The concept self-organized criticality (SOC) has attracted considerable interest in many areas of science  including ﬁelds in pure mathematics influenced by string theory \cite{kalinin2018self}.

A procedure of identification of SOC systems is to measure the distribution in space and time of events in the system’s evolution, and seek for power law behavior. Then, self-organized criticality can be identified when the distribution of the sizes in a cascade of events (also called an avalanche) follows a power law

\begin{eqnarray}
P \left( s \right) = s^{- \tau},    
\end{eqnarray}

\noindent where the exponent $\tau$, $s$ and $P \left( s \right)$ are a positive constant also called the Fisher exponent \cite{Fisher:1967dta}, the size of an avalanche, and the distribution of the size of the avalanche, respectively. The power-law distribution indicates that there is no preferred scale characterizing the avalanches, implying that the system exhibits the same structure over all scales. This remarkable feature is typical of physical systems where a fragmentation process producing a cluster distribution of objects is observed. Examples of systems exhibiting such fragmentation and clusterization phenomena are nuclear fragment distributions following a collision of two heavy ions \cite{finn1982nuclear}, avalanche \cite{kadanoff1989scaling} and sandpile models\cite{bak1987self,bak1988self} or percolation cluster distributions \cite{stauffer2018introduction}. 

From the above statistical fragmentation model, we want to show that the log sector evolves as a model of fragmentation and partitioning based on the symmetric group $S_n$ that leads to a power law behavior in the cluster size distribution function. We first note that when $x=1$, the Gibbs distribution in Eq. (\ref{Gibbs distribution}) becomes the Hurwitz number in Eq. (\ref{Hurwitz numbers}). From there, we establish the following.

\paragraph{Theorem.} \textit{The average number of clusters of size $k$ in the fragmentation process taking place in the logarithmic sector of cosmological topologically massive gravity at the critical point is}

\begin{eqnarray}
\label{1/k}
\langle j_k \rangle= \sum_{\vec{j}_k} j_k \cdot H^{\bullet}_{0 \xrightarrow[]{n} 0} \left[ \left( [k]^{j_k} \right)_{k=1}^n,  \left( [k]^{j_k} \right)_{k=1}^n \right]= \frac{1}{k}.
\end{eqnarray}

\noindent \textit{Proof.} In order to prove the above theorem, we to simply count all distinct pairs of a permutation $\pi$ of $S$, and a specific $k$-cycle (a set of size $k$) that occurs as orbit for $\pi$; this number should turn out to be $n!/k$ which then proves that the expected number of $k$-cycles associated to a random $\pi$ is $1/k$. Let $n \geq k$. Working in the symmetric group $S_n$  of permutations of $[n]=\left\{ 1,2, \ldots,n \right\}$, we start by enumerating the possible $k$-subsets of $S$, yielding $\left(\begin{smallmatrix}n \\ k\end{smallmatrix}\right)$ as possibilities for orbits. The first remark is that, given a fixed $k$-cycle, it appears in exactly $(n-k)!$ elements of $S_n$. (This proceeds from all the permutations of the remaining $n-k$ elements of $[n]$.) The second remark is that, there are $\left(\begin{smallmatrix}n \\ k\end{smallmatrix}\right) \cdot \frac{k!}{k}$ cycles of length $k$ that can be made with elements of $[n]$. (Choose $k$ elements of $[n]$ to form a $k$-cycle , and then choose a cyclic permutation of these elements.) Counting over all of $S_n$, there will be a total of $(n-k)! \left(\begin{smallmatrix}n \\ k\end{smallmatrix}\right) \cdot \frac{k!}{k}= \frac{n!}{k}$ cycles of length $k$ appearing in all of $S_n$. Eventually, the expected number of cycles of length $k$ appearing in all of $S_n$ is given by $\frac{\frac{n!}{k}}{n!}$. Note that the appearance of the result $\frac{1}{k}$ is due to the fact that a cycle of length $k$ can be expressed in $k$ different ways. $\blacksquare$

The above theorem connecting equivalently cluster size distribution or expected number of $k$-cycles to the double Hurwitz number tells us that the ensemble averaging over all partitions leads to a scale invariant power law behavior, as $\langle j_k \rangle$ does not depend on $n$. The cluster size distribution given in terms of the mean (ensemble averaged) number of clusters $\langle j_k \rangle$ as a function of the size $k$ suggests that our result in Eq.(\ref{1/k} ) is functionally the same as the sandpile distribution function $D(s)$ in \cite{bak1987self}, where $s$ is the size of the slide, and $D(s) \sim s^{-1}$ at the self-organized critical point.

\section{Fragmentation classification of disorder in log gravity}
The purpose of this section is to present a similarity between spin glass models \cite{castellani2005spin}, one of the known members of the SOC family \cite{le2010avalanches}, and log gravity. The motivation for this is to show that they both can be contained in a unified way in a same simple expression under which various disordered systems fall, allowing for a probabilistic classification of the disorder in log gravity in terms of random fragmentation.

The breakup of the log $n$-particle sector into clusters of various sizes has been shown to behave according to the Ewens fragmentation pattern. Various physical systems also exhibit clustering. In condensed matter physics for instance, the phase space of disordered systems \cite{derrida1987statistical,ziman1979models} such as spin glasses \cite{mezard1987spin,chowdhury2014spin} exhibits clustering of states, and the nature of the disorder can be better understood by studying the sizes and distributions of the clusters. 

The log sector of CCTMG can be shown to belongs to a class of disordered systems by pointing out that the random breaking of collective excitations in the log $n$-particle sector is part of a family of random fragmentation processes in disordered systems that share a particular probability distribution given by

\begin{eqnarray}
\label{betabin}
P_r (N, x, \gamma) = \begin{pmatrix} N \\ r \end{pmatrix}  \frac{\rm{Beta} \left[ N - r + x, r + \gamma \right]}{\rm{Beta} \left[ x, \gamma \right]}, 
\end{eqnarray}

\noindent where $x$ and $\gamma$ are parameters, $r$ is a random variable which can take values $r=0,1,\ldots,N$, and the beta functions are expressed in terms of gamma functions as

\begin{eqnarray}
\label{betaprop1}
\rm{Beta} \left[ a,b \right] = \frac{\Gamma (a) \Gamma (b)}{\Gamma (a+b)}.   
\end{eqnarray}

\noindent In probability theory, $P_k (N, x, \gamma)$ is known as a Poly\'{a} distribution \cite{feller1971introduction}. Table \ref{table 1} whose values for $x$ and $\gamma$ are taken from \cite{mekjian1999disoriented} shows the Poly\'{a} distribution classification of disorder in spin glass and log gravity.

\begin{table}[h]
\begin{center}
\renewcommand*{\arraystretch}{1.8}
\begin{tabular}{|c|ccc|c|} 
\hline 
Physical model & $x$ & & $\gamma$ & Quantity described \\
\hline \hline
Spin glass \cite{mezard1987spin} & & $x+ \gamma = 1$ & & Well depth from random Hamiltonians \\
\hline
Log gravity &1& &1& Cycle length\\
\hline
\end{tabular}
\caption{\label{table 1} Poly\'{a} distribution classification of disorder in spin glass and log gravity}
\end{center}
\end{table}

\noindent The above statements are elaborated in the three subsections below as follows. We first give a brief mathematical background on Eq. (\ref{betabin}). We thereafter give an presentation of its application to log gravity. Finally, we conclude by showing how the probability distribution arises in spin glasses and can be used to bring log gravity and spin glasses in a particular class of disordered systems.

\subsection{Mathematical background}
We begin with the notion of conditional probability, which is a basic tool in probability theory \cite{feller1971introduction,dasgupta2010fundamentals}, and give a Bayesian derivation of $P_r (N, x, \gamma)$.

Let $B$ be an event with positive probability. For an arbitrary event $A$, the quantity 

\begin{eqnarray}
P (A|B)= \frac{P(AB)}{P(B)}    
\end{eqnarray}

\noindent is the conditional probability of $A$ on the hypothesis (or for given) $B$. If we consider $N$ independent Bernoulli trials (0-1 trials) with probability parameter $p$, it is known that the number of successes $r$ has the binomial distribution. Assuming the probability parameter $p$ unknown (with the sample-size parameter $N$ known), we have $r|p \sim \text{bin} (N,p)$, and the generic expression

\begin{eqnarray}
f (r|p)= \begin{pmatrix} N \\ r \end{pmatrix} p^r (1-p)^{1-r}, \hspace{1cm} r=0,1,\ldots,N.    
\end{eqnarray}

\noindent Taking as prior distribution for $p$ a beta distribution, i.e. $p \sim \text{beta} (\gamma,x)$, with density function

\begin{eqnarray}
\label{fp}
f(p)= \frac{1}{\text{Beta} (\gamma,x)} p^{\gamma-1} (1-p)^{x -1}, \hspace{1cm} 0< p< 1,   
\end{eqnarray}

\noindent the posterior distribution is also a beta distribution, i.e $p|r \sim \text{beta} (\gamma +r,x + N - r)$, and we have the simultaneous distribution 

\begin{eqnarray}
f(r,p) = f(p) f(r|p) = \frac{\begin{pmatrix} N \\ r \end{pmatrix}}{\text{Beta} (\gamma,x)} p^{\gamma+r-1} (1-p)^{x + N -r -1}.    
\end{eqnarray}

\noindent Then, integrating $p$ out, we get 

\begin{subequations}
\begin{align}
P_r(N,x,\gamma) =& \begin{pmatrix} N \\ r \end{pmatrix} \int_0^1 f(r|p) f(p) dp, \hspace{2cm}  \text{with} \quad \int_0^1 f(p) dp, \label{betabin1} \\ 
=& \begin{pmatrix} N \\ r \end{pmatrix} \frac{\text{Beta} (\gamma +r,x + N - r)}{\text{Beta} (\gamma,x)},  \hspace{1cm} r=0,1,\ldots,N, \label{betabin2}
\end{align}    
\end{subequations}

\noindent which is also called the beta-binomial distribution, i.e. $r \sim \text{betabin} (N,x,\gamma)$. Using the beta function property in Eq. (\ref{betaprop1}) and the gamma function property

\begin{eqnarray}
 \Gamma (\gamma + r) = \Gamma (\gamma ) \gamma (\gamma +1) \ldots   (\gamma + r-1),
\end{eqnarray}

\noindent we also obtain 

\begin{eqnarray}
\label{betabin3}
P_r(N,x,\gamma) = \begin{pmatrix} N \\ r \end{pmatrix} \frac{\gamma (\gamma +1) \ldots   (\gamma + r-1) x (x +1) \ldots   (x + N-r-1)}{(\gamma +x)(\gamma +x+1)\ldots (\gamma +x+N-1)}.   
\end{eqnarray}

Finally, we mention that when $\gamma$ and $x$ are integers, Eq. (\ref{betabin3}) gives rise to interpreting the beta-binomial distribution by an urn model. Because the Hungarian mathematician P\'{o}lya derived the beta-binomial distribution in this way \cite{eggenberger1923statistik}, it is also called the P\'{o}lya distribution.

\subsection{P\'{o}lya distribution in log gravity}
We next show how the P\'{o}lya distribution applies to the case of log gravity. The log partition function generates counts of random permutations weighted by the number of cycles, where a permutation $\pi \in S_n$ of $\left\{ 1, \ldots, n \right\}$ is decomposed as a product of cycles with $\pi$ chosen uniformly with probability $1/n!$ and distributed according to the Ewens sampling formula, by encoding a permutation of $n$ with exactly $j_k$ cycles of length (or size) $k$ in the product

\begin{eqnarray}
\label{prod}
\prod_{k=1}^n \mathcal{G}_k^{j_k}.    
\end{eqnarray}

The information captured by Eq. (\ref{prod}) can be associated to the elements (balls) of an urn as follows. Consider an urn that consists of elements $\left\{ \textcolor{cyan}{\bullet}  ~ \textcolor{orange}{\bullet} ~ \textcolor{magenta}{\bullet} ~ \textcolor{cyan}{\bullet} ~ \textcolor{magenta}{\bullet} ~ \textcolor{cyan}{\bullet} ~ \textcolor{cyan}{\bullet} ~ \textcolor{orange}{\bullet} \right\}$ of length (or size) $n=8$. If we arrange the colors in clusters, and denote the clusters with ket notation as states, the states $\ket{\textcolor{cyan}{\bullet}   \textcolor{cyan}{\bullet}  \textcolor{cyan}{\bullet}  \textcolor{cyan}{\bullet}}$, $\ket{\textcolor{magenta}{\bullet} \textcolor{magenta}{\bullet}}$ and $\ket{\textcolor{orange}{\bullet} \textcolor{orange}{\bullet}}$ each appearing with multiplicity one can be encoded as $\mathcal{G}^{\textcolor{cyan}{1}}_{\textcolor{cyan}{\bullet} \textcolor{cyan}{\bullet} \textcolor{cyan}{\bullet} \textcolor{cyan}{\bullet}} \mathcal{G}^{\textcolor{magenta}{1}}_{\textcolor{magenta}{\bullet} \textcolor{magenta}{\bullet}} \mathcal{G}^{\textcolor{orange}{1}}_{\textcolor{orange}{\bullet} \textcolor{orange}{\bullet}}$. Finally, if we are only interested in the cluster size (i.e, the cycle length), we have the equivalence

\begin{eqnarray}
\mathcal{G}^{\textcolor{cyan}{1}}_{\textcolor{cyan}{\bullet} \textcolor{cyan}{\bullet} \textcolor{cyan}{\bullet} \textcolor{cyan}{\bullet}} \mathcal{G}^{\textcolor{magenta}{1}}_{\textcolor{magenta}{\bullet} \textcolor{magenta}{\bullet}} \mathcal{G}^{\textcolor{orange}{1}}_{\textcolor{orange}{\bullet} \textcolor{orange}{\bullet}}  \equiv \mathcal{G}^1_4 \mathcal{G}^1_2 \mathcal{G}^1_2 = \mathcal{G}^1_4 \mathcal{G}^2_2,   
\end{eqnarray}

\noindent where the partition of $n=8$ appears as $1 \cdot 4 + 2 \cdot 2$. This equivalence enables us to proceed with the description of the log sector as an urn model.

The $n$-particle log sector is then described in terms of the Hoppe urn model as follow. The Hoppe urn initially contains only one black ball, and at each $n$-particle level contains one black ball and $n$ colored balls. Starting at the single particle level, one ball is randomly drawn from the urn. If the black ball is drawn, it is reintroduced in the urn together with a new ball of a new color. If instead, a colored ball is drawn, it is reintroduced in the urn together with a new ball of the same color, as in the standard P\'{o}lya urn of interest here. The Hoppe's urn model which  shows that the color composition $(j_1, \ldots, j_n)$ is distributed according to the Ewens sampling formula, where $j_k$ is the number of colors represented by exactly $k$ of the first $n$ non-black balls, with $1 \leq k \leq n$, offers an alternative and colorful way of viewing the instability in the log sector of log gravity. Indeed, the dynamic instability is marked by the appearance of new colors when running the Hoppe's urn model forward in time (i.e as the $n$-particle system evolves).

The simplest way to obtain the P\'{o}lya distribution within the Hoppe's urn model is from the following general random process. Starting with $n_{\ket{\textcolor{cyan}{\bullet}}}$ cyan balls and $n_{\ket{\textcolor{black}{\bullet}}}$ black balls, a ball is drawn at random and replaced with $n_{\ket{+}}$ additional balls of the same color. The probability of drawing $r$ cyan balls in $N$ trials is then given by the P\'{o}lya distribution where $\gamma = n_{\ket{\textcolor{cyan}{\bullet}}}/ n_{\ket{+}}$ and $x = n_{\ket{\textcolor{black}{\bullet}}}/ n_{\ket{+}}$. Since we start at the $1$-particle level where $n_{\ket{\textcolor{cyan}{\bullet}}}=n_{\ket{\textcolor{black}{\bullet}}}=1$, and we are interested in the addition of only one cyan ball, we have $\gamma = x= 1$. Furthermore, we are looking at the probability of drawing 1 cyan ball in 1 trial, 2 cyan balls in 2 trials, 3 cyan balls in 3 trials, etc... It is well known (\cite{dasgupta2010fundamentals} page 392), that the P\'{o}lya distribution in this case reduces to 

\begin{eqnarray}
\label{polyahurwitz}
 P_{r=N}(N,1,1) = \begin{pmatrix} N \\ N \end{pmatrix} \frac{N!(N-r)!}{2\times 3 \times \cdots \times (N+1)} = \frac{1}{N+1}. 
\end{eqnarray}

A simple way of illustrating the P\'{o}lya distribution in log gravity up to $3$-particle level, i.e for $N=2$ draws using Eq. (\ref{betabin3}) is to realize that starting at the $1$-particle level again, if we first draw a blue ball, put the ball back into the urn with another blue ball and then draw another blue, the probability in the case $\left\{ (r=N;N,x,\gamma)= (2;2,x,\gamma) \right\}$ given from Eq. (\ref{betabin3}) as

\begin{eqnarray}
P_{r=2}(2,x,\gamma)=  \begin{pmatrix} 2 \\ 2 \end{pmatrix} \frac{\gamma  }{(\gamma +x)} \cdot \frac{(\gamma +1)}{(\gamma +x+1)},   
\end{eqnarray}

\noindent becomes

\begin{eqnarray}
P_{r=2}(2,1,1)= P \left( \text{Cyan}_1 \text{Cyan}_2 \right) =  P \left( \text{Cyan}_1 \right) \cdot P \left( \text{Cyan}_2 | \text{Cyan}_1 \right) = \frac{1}{2}  \cdot \frac{2}{3} = \frac{1}{3}.
\end{eqnarray}

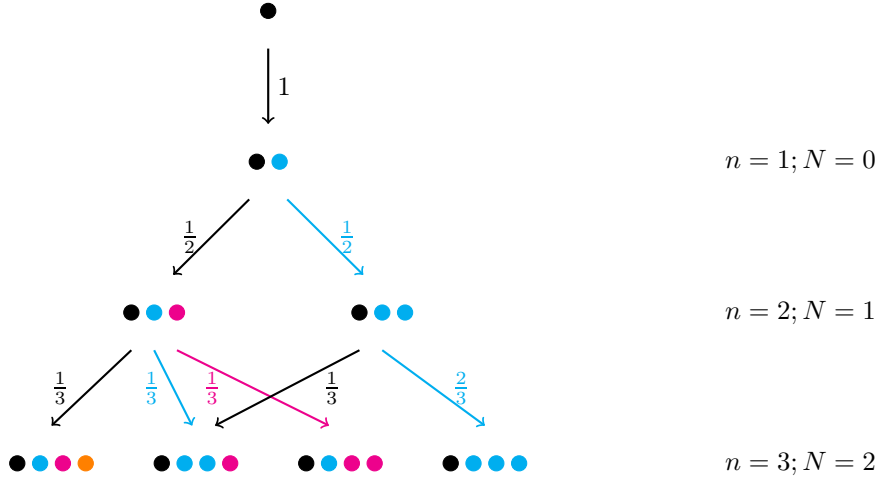
\begin{figure}[h]
\begin{center} 
\begin{tikzpicture}
\filldraw[black] (0,0) circle (0.1);
\draw[->,thick] (0,-0.5) -- (0,-1.5);
\node [right] at (0,-1) {1};
\filldraw[black] (-0.15,-2) circle (0.1);
\filldraw[cyan] (+0.15,-2) circle (0.1);
\draw[->,thick] (-0.25,-2.5) -- (-1.25,-3.5);
\node [left] at (-0.8,-3) {$\frac{1}{2}$};

\draw[->,thick,cyan] (+0.25,-2.5) -- (+1.25,-3.5);
\node [right,cyan] at (+0.8,-3) {$\frac{1}{2}$};
\filldraw[black] (-1.8,-4) circle (0.1);
\filldraw[cyan] (-1.5,-4) circle (0.1);
\filldraw[magenta] (-1.2,-4) circle (0.1);

\filldraw[black] (+1.2,-4) circle (0.1);
\filldraw[cyan] (+1.5,-4) circle (0.1);
\filldraw[cyan] (+1.8,-4) circle (0.1);

\draw[->,thick] (-1.8,-4.5) -- (-2.85,-5.5);
\node [left] at (-2.5,-5) {$\frac{1}{3}$};

\draw[->,thick,cyan] (-1.5,-4.5) -- (-1,-5.5);
\node [left,cyan] at (-1.3,-5) {$\frac{1}{3}$};

\draw[->,thick,magenta] (-1.2,-4.5) -- (+0.8,-5.5);
\node [left,magenta] at (-0.5,-5) {$\frac{1}{3}$};

\draw[->,thick] (+1.2,-4.5) -- (-0.7,-5.5);
\node [right] at (+0.6,-5) {$\frac{1}{3}$};

\draw[->,thick,cyan] (+1.5,-4.5) -- (+2.85,-5.5);
\node [right,cyan] at (+2.3,-5) {$\frac{2}{3}$};


\filldraw[black] (-3.3,-6) circle (0.1);
\filldraw[cyan] (-3,-6) circle (0.1);
\filldraw[magenta] (-2.7,-6) circle (0.1);
\filldraw[orange] (-2.4,-6) circle (0.1);

\filldraw[black] (-1.4,-6) circle (0.1);
\filldraw[cyan] (-1.1,-6) circle (0.1);
\filldraw[cyan] (-0.8,-6) circle (0.1);
\filldraw[magenta] (-0.5,-6) circle (0.1);

\filldraw[black] (+0.5,-6) circle (0.1);
\filldraw[cyan] (+0.8,-6) circle (0.1);
\filldraw[magenta] (+1.1,-6) circle (0.1);
\filldraw[magenta] (+1.4,-6) circle (0.1);

\filldraw[black] (+2.4,-6) circle (0.1);
\filldraw[cyan] (+2.7,-6) circle (0.1);
\filldraw[cyan] (+3,-6) circle (0.1);
\filldraw[cyan] (+3.3,-6) circle (0.1);

\node [] at (+7,-2) {$n=1;N=0$};

\node [] at (+7,-4) {$n=2;N=1$};

\node [] at (+7,-6) {$n=3;N=2$};
\end{tikzpicture}    
\end{center}
\caption{Log sector evolution up to 3-particle level in terms of sample paths for $n=3$ draws from Hoppe’s urn. $N$ denotes the number of draws in the P\'{o}lya process considered.}
\label{fig1}
\end{figure}

\noindent The above example can be visualized in the Hoppe’s urn construction drawn in Fig. (\ref{fig1}), which can also be found in \cite{mahmoud2008polya}. Note that the arrows in the diagram are labeled with a corresponding probability; an arrow coming down from one configuration into another is labeled with the probability of making this particular transition, and the color of the ball drawn in the immediate anterior configuration.

The log sector evolution in Fig. (\ref{fig1}) explicitly shows the branch of the hierarchical sequence of events in which the P\'{o}lya process considered takes place. Note that the number of draws $N$ in the P\'{o}lya process under consideration within the Hoppe model is a unit less than the number of draws $n$ from the overall Hoppe scheme at the $n$-particle level.

\subsection{P\'{o}lya distribution in spin glasses}
We conclude this section with an analysis largely inspired from results obtained in the study of fragmentation of nuclei \cite{chase1998randomly}. The intention is, by giving a probabilistic description of the similarity between log gravity and spin glasses, to present the two theories as models that belong to a class of disordered systems related to each other by the P\'{o}lya distribution.   

When $P_r(N,x,\gamma)$ is given as in Eq. (\ref{betabin1}), the important function $f(p)$ in Eq. (\ref{fp}). emerges. The latter function which we relabel as $f(p,x,\gamma)$ and its associated function $\mathcal{F}(p,x,\gamma)= p^{-1} f(p,x,\gamma)$ play an important role in the theory of disordered systems. In spin-glass models, random Hamiltonians based on an Ising interaction $J \vec{\sigma}_i \vec{\sigma}_j$ (where $J=\pm \left|j\right|$ is chosen randomly) are used to calculate rugged free-energy landscapes. The function $\mathcal{F}(p,x,\gamma)$ and its related function
$f(p,x,\gamma)$ give the distribution of well depths in the free energy landscape.

We turn to the many interesting properties of the quantity $\langle j_k \rangle$ for clusters and breaking processes with size $n$ into fragments of size $k=1,2,\ldots,n$ with $j_k$ as the number of fragments of size $k$ under the constraint given by Eq. (\ref{constraint}). An important feature in the present work is the appearance of a power law in the fragment distribution $\langle j_k \rangle = 1/k$, when $x=\gamma=1$. This power law can be approximated in the more general case at large $n$ by 

\begin{eqnarray}
\label{discret}
\langle j_k \rangle \simeq \frac{\Gamma (x+\gamma)}{\Gamma(x)\Gamma(\gamma)} \frac{1}{k} \left(1-\frac{k}{n}\right)^{x-1} \left(\frac{k}{n}\right)^{\gamma-1}.   
\end{eqnarray}

\noindent When $x=1$, $\langle j_k \rangle \simeq \gamma n^{1-\gamma} k^{\gamma-1}$, which shows a power law with Fisher exponent $\tau=2-\gamma$. When $\gamma=1$, $\langle j_k \rangle \simeq (x/k) \left[ 1-(k/n)\right]^{x-1}$, which is the behavior of $\langle j_k \rangle$ obtained in a model
based on the sequential partitioning of a system of size $n$ into  subsystems of sizes $k_1,k_2,\ldots$ \cite{frontera1995sequential}, when relating our $x$ parameter with their $\beta$ parameter by $\beta=x-1$. The latter model was proposed as a useful description of avalanche size distributions in first order phase transitions in several physical systems, and is the discrete case of the random breaking of an interval model which we now discuss.

The above expression of $\langle j_k \rangle$ corresponds to a discrete fragmentation, but can be turned into a continuum model by setting $n \longrightarrow \infty$ and $k/n \longrightarrow W$. In this limit, the discrete distribution $\langle j_k \rangle$ given by Eq. (\ref{discret}) becomes a continuous distribution $\mathcal{F}(W)= \lim_{n \rightarrow \infty} \langle j_{nW} \rangle$ which satisfies $\int_0^1 W \mathcal{F}(W) =1$ and is given by

\begin{eqnarray}
\label{dfdisor}    
\mathcal{F}(W,x,\gamma)= \frac{\Gamma (x+\gamma)}{\Gamma(x)\Gamma(\gamma)} \left(1-W\right)^{x-1} \left(W\right)^{\gamma-2},   
\end{eqnarray}

\noindent where $W\mathcal{F}(W)$ is a beta distribution. Eq. (\ref{dfdisor}) has appeared in contexts related to disordered systems \cite{derrida1987statistical,derrida1986multivalley,derrida1987random}, extending previous results on spin glasses \cite{mezard1984replica}. The cases considered in these models, introduced to study disordered systems in which phase space is split into many valleys or basins of attraction having different weights or sizes, $W_s$ being the weight of the $s$-valley, are in fact special cases of Eq. (\ref{dfdisor}). In the particular case $\gamma=y, x=1-y$, Eq. (\ref{dfdisor}) reduces to the spin glass $f(W)$ in \cite{derrida1987statistical} (see Eqs. (11a) and (41b) therein), reproducing the Sherrington-Kirkpatrick spin-glass model \cite{mezard1984replica}, where $y$ contains the physical quantities such as temperature and spin coupling strength.

\subsection{Feynman diagram representation in  spin glasses and log gravity}
Although as seen in the above subsections, the P\'{o}lya distribution originates as a distribution for a replacement urn model, its relevance in classifying (randomly broken) disordered systems takes a particular meaning as we consider systems with disorder defined on graphs, of which spin glasses are a well-known example. Below, we bring further arguments to show that log gravity falls within this category, exhibiting disorder on trees.

Disordered systems are usually defined on random graphs and rooted trees \cite{cugliandolo2019advanced}. As mentioned in the previous subsection, the $x+\gamma=1$ special case of the P\'{o}lya distribution arrived at in \cite{derrida1987statistical} reproduces the Sherrington-Kirkpatrick spin-glass model with quenched disorder \cite{castellani2005spin}. The latter is defined on a complete graph \cite{panchenko2014introduction,cugliandolo2013out}. For several decades, there has been a strong interest in the statistical investigation of spin glasses on random graphs of various types. In particular, the investigation of spin models, especially spin glasses, on Feynman diagrams using methods borrowed from the fat graphs of two-dimensional gravity has been discussed in the literature. Indeed, a series of papers \cite{Baillie:1994wd,Baillie:1995ek,Baillie:1995wi,baillie1996dynamic,Baillie:1995in} ensued after a refined method of describing spin models on random graphs, inspired from the matrix model methods used to describe planar random graphs in two-dimensional gravity, was proposed in \cite{Bachas:1994qn}. The idea in the proposal for looking at the problem of spin models living on random graphs was to observe that the requisite ensemble of random graphs could be generated by considering the Feynman diagram expansion for the partition function of the model. The result of this train of thoughts in the investigation of the behaviour of various spin models on Feynman diagrams is that quantum gravity graphs provided a rich laboratory for analytical and numerical investigations of disordered statistical physics systems \cite{janke2006two}. More recently, counting procedures of metastable states of certain spin glasses on connected Feynman diagrams expansions summed by the logarithm of the partition function have been performed \cite{waclaw2008counting}. In a similar manner, we show below that the disorder in log gravity can be defined on connected Feynman diagrams, which are the ladder rooted trees summed in the logarithm of the partition function indexed by the connected Hurwitz number studied in \cite{Mvondo-She:2022jnf}. Hurwitz numbers are topological invariants indexed by tuples of partitions that count branched coverings of two-dimensional Riemann surfaces. These enumerative objects also have many other interpretations such as coefficients in symmetric groups, numbers enumerating certain classes of graphs or again numbers generated by functions which provide solutions to integrable Kadomtsev-Petviashvili (KP) and Toda hierarchies, all which have appeared in the study of log gravity.
A fundamental result in direction of a probabilistic interpretation of Hurwitz numbers will arise through an unexpected connection between the connected Hurwitz number studied in \cite{Mvondo-She:2022jnf} and the P\'{o}lya distribution $ P_{r=N}(N,1,1)$ in Eq. (\ref{polyahurwitz}).

We start by giving a genus-zero Feynman diagram interpretation of Hoppe's urn model from the latter's description of the log sector. A family of random tree model associated to the Hoppe’s urn model has appeared in the literature under the name of Hoppe tree \cite{leckey2013asymptotic}. In the special case $\theta=1$ which is ours, this tree model is also known as the random recursive tree model, in which the recursive trees grow in a similar manner as permutations in cycle notation. Given the structure of the log partition function which can be understood in terms of the distribution of cycle counts according to the $\theta$-biased permutation with parameter $\theta=1$, the content of the urns in Hoppe's scheme corresponds to the graphical representation of the log partition function in terms of genus-zero Feynman diagrams \cite{Mvondo-She:2022jnf}. The $n$-level Feynman diagram dynamics form a randomly growing rooted tree model whose growth parallels the evolution of Hoppe's urn model.

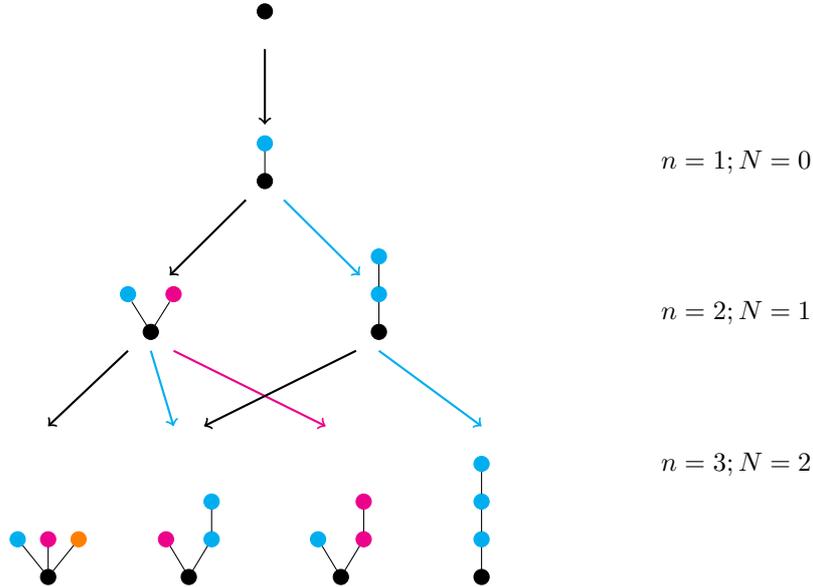
\begin{figure}[h]
\begin{center} 
\begin{tikzpicture}
\filldraw[black] (0,0) circle (0.1);
\draw[->,thick] (0,-0.5) -- (0,-1.5);
\filldraw[black] (0,-2.25) circle (0.1);
\draw[-] (0,-1.75) -- (0,-2.25);
\filldraw[cyan] (0,-1.75) circle (0.1);
\draw[->,thick] (-0.25,-2.5) -- (-1.25,-3.5);

\draw[->,thick,cyan] (+0.25,-2.5) -- (+1.25,-3.5);
\filldraw[cyan] (-1.8,-3.75) circle (0.1);
\draw[-] (-1.5,-4.25) -- (-1.75,-3.85);
\filldraw[black] (-1.5,-4.25) circle (0.1);
\draw[-] (-1.5,-4.25) -- (-1.25,-3.85);
\filldraw[magenta] (-1.2,-3.75) circle (0.1);

\filldraw[black] (+1.5,-4.25) circle (0.1);
\draw[-] (+1.5,-4.25) -- (+1.5,-3.75);
\filldraw[cyan] (+1.5,-3.75) circle (0.1);
\draw[-] (+1.5,-3.65) -- (+1.5,-3.25);
\filldraw[cyan] (+1.5,-3.25) circle (0.1);

\draw[->,thick] (-1.8,-4.5) -- (-2.85,-5.5);

\draw[->,thick,cyan] (-1.5,-4.5) -- (-1.2,-5.5);

\draw[->,thick,magenta] (-1.2,-4.5) -- (+0.8,-5.5);

\draw[->,thick] (+1.2,-4.5) -- (-0.8,-5.5);

\draw[->,thick,cyan] (+1.5,-4.5) -- (+2.85,-5.5);
\filldraw[black] (-2.85,-7.5) circle (0.1);
\draw[-] (-2.85,-7.5) -- (-3.25,-7);
\filldraw[cyan] (-3.25,-7) circle (0.1);
\draw[-] (-2.85,-7.5) -- (-2.85,-7);
\filldraw[magenta] (-2.85,-7) circle (0.1);
\draw[-] (-2.85,-7.5) -- (-2.45,-7);
\filldraw[orange] (-2.45,-7) circle (0.1);

\filldraw[black] (-1.0,-7.5) circle (0.1);
\draw[-] (-1.0,-7.5) -- (-1.3,-7);
\filldraw[magenta] (-1.3,-7) circle (0.1);
\draw[-] (-1.0,-7.5) -- (-0.7,-7);
\filldraw[cyan] (-0.7,-7) circle (0.1);
\draw[-] (-0.7,-6.9) -- (-0.7,-6.5);
\filldraw[cyan] (-0.7,-6.5) circle (0.1);

\filldraw[black] (+1.0,-7.5) circle (0.1);
\draw[-] (+1.0,-7.5) -- (+0.7,-7);
\filldraw[cyan] (+0.7,-7) circle (0.1);
\draw[-] (+1.0,-7.5) -- (+1.3,-7);
\filldraw[magenta] (+1.3,-7) circle (0.1);
\draw[-] (+1.3,-6.9) -- (+1.3,-6.5);
\filldraw[magenta] (+1.3,-6.5) circle (0.1);

\filldraw[black] (+2.85,-7.5) circle (0.1);
\draw[-] (+2.85,-7.5) -- (+2.85,-7.0);
\filldraw[cyan] (+2.85,-7.0) circle (0.1);
\draw[-] (+2.85,-6.9) -- (+2.85,-6.5);
\filldraw[cyan] (+2.85,-6.5) circle (0.1);
\draw[-] (+2.85,-6.4) -- (+2.85,-6.0);
\filldraw[cyan] (+2.85,-6.0) circle (0.1);

\node [] at (+6.2,-2) {$n=1;N=0$};

\node [] at (+6.2,-4) {$n=2;N=1$};

\node [] at (+6.2,-6) {$n=3;N=2$};
\end{tikzpicture}    
\end{center}
\caption{Log sector evolution in terms of sample paths in Hoppe (rooted) trees as Feynman diagrams up to order $n=3$.}
\label{fig3}
\end{figure}

We previously argued \cite{Mvondo-She:2023xel} that in the log partition function expansion over rooted trees, solitonic clusters are pinned at the roots which represent the pinning sites. The interesting observation here is that, as the balls in the Hoppe urns are represented by nodes in the Hoppe trees, the mutator ball that initiates the Hoppe's urn model corresponds to the root of the Hoppe tree, responsible for the diffusion of the fragmented clusters that realize the disorder landscape (see Fig. \ref{fig3}). 

Recalling the definition of Hurwitz numbers \cite{cavalieri2016riemann}, let $Y$ be a connected Riemann surface of genus $g$. Define the set $B = \{ y_1, \ldots, y_d \} \in Y $, and let $\lambda_1, \ldots, \lambda_d$ be partitions of the positive integer $n$. Then the Hurwitz number can be defined as the sum

\begin{eqnarray}
H_{X \xrightarrow[]{n} Y} \left( \lambda_1, \ldots, \lambda_d \right) = \sum_{\left|f \right|} \frac{1}{\left| \rm{Aut} (f) \right|}
\end{eqnarray}

\noindent that runs over each isomorphism class of $f: X \mapsto Y$ where

\begin{enumerate}
    \item $f$ is a holomorphic map of Riemann surfaces;
    \item $X$ is connected and has genus $h$;
    \item the branch locus of $f$ is $B = \{ y_1, \ldots, y_d \}$;
    \item the ramification profile of $f$ at $y_i$ is $\lambda_i$.
\end{enumerate}

\noindent After restricting the genus of the base and target Riemann surfaces to be zero, then further imposing $d=2$ and $\lambda_1 = \lambda_2 = (n)$, the expression of connected Hurwitz numbers becomes

\begin{eqnarray}
\label{connected HN}
H_{0 \xrightarrow[]{n} 0} \left( (n), (n) \right) = \frac{1}{n}.
\end{eqnarray}

A fundamental result appears when we bring Eq. (\ref{connected HN}) in contact with Eq. (\ref{polyahurwitz}). Indeed, as mentioned earlier, the number of draws $N$ in the P\'{o}lya process under consideration within the Hoppe model is a unit less than the number of draws $n$ from the overall Hoppe scheme at the $n$-particle level, i.e $n=N+1$. We then have 

\begin{eqnarray}
H_{0 \xrightarrow[]{n} 0} \left( (n), (n) \right) =  P_{r=N}(N,1,1), 
\end{eqnarray}

\noindent and we rewrite the log partition function in terms of the P\'{o}lya distribution $P_{r=N}(N,1,1)$ as

\begin{eqnarray}
Z_{log} \left( q, \bar{q} \right) = \exp(\sum_{n=1}^\infty \left[ P_{r=N}(N,1,1) \right] \cdot \mathcal{G}_n \left( q, \bar{q} \right) \cdot \left( q^2 \right)^n).  
\end{eqnarray}

It is interesting to observe that the disconnected and the connected Hurwitz numbers both have a probabilistic meaning in this theory. While the disconnected Hurwitz number is identical to the $\theta=1$ Ewens distribution, its connected counterpart is the P\'{o}lya distribution $P_{r=N}(N,1,1)$.

Through the analogy with spin glass models which are an important example of (randomly broken) disordered system on graphs in which features the P\'{o}lya distribution, we have from a probabilistic perspective shown how the disorder appears in log gravity, by simply considering how the fragmented linear branch of our Feynman diagrams is indexed according to the P\'{o}lya distribution. Log gravity then appears as an instance of a system with disorder on Feynman diagrams.

An interest was raised in \cite{derrida1988polymers}, to know whether the analogy between the mean field theory of spin glasses and the theory of traveling waves could be pushed further, after showing that the problem of a directed polymer on a tree with disorder could be reduced to the study of nonlinear equations that admit traveling wave solutions. It is our belief that our works on log gravity bring a positive response to the interest.

\section{Summary and outlook}
In this paper, we have used the canonical Gibbs model as a statistical framework to discuss a nonequilibrium fragmentation phenomenon in the partition-valued stochastic dynamics of the log sector in log gravity. 

In this description, a statistical Gibbs weight is given to each partition of $n$ into clusters of size $k$, which correspond to cycles of size $k$ in the symmetric group. Summing over all partitions with constraint (\ref{constraint}) leads to a probability distribution representing the probability of an element being in a cluster of size $k$, that is used to study the distribution of cluster sizes by ensemble averaging. 

We prove that the ensemble averaged distribution exhibits a scale-invariant power-law behavior. The fact that the expected number of clusters of size $k$ functionally behaves as an inverse power of the cluster size $k$ is a signature of self-organised  criticality, a specific type of non-equilibrium collective behaviour associated with a critical point. 

The stochastic dynamics in the log sector was previously shown to describe a non-equilibrium process displaying aging behavior. Motivated by this glassy dynamics resemblance, a similarity between spin glass models and log gravity is pointed out. Not only they both are members of the self-organized criticality family, but it appears that they also belong to a family of disordered systems classified by the P\'{o}lya distribution. By showing how the log partition function can be expressed in terms of a limit of the P\'{o}lya distribution, as well as showing how the connected Hurwitz number plays the role of the distribution, a disorder on Feynman diagrams has been discussed. 

From our study of log gravity, it is observed that at the critical point $\mu l =1$, the log sector shows a remarkable dynamical self-organization into a solitonic configuration that forms a KP hierarchy. The sandpile model being a paradigmatic example of self-organized criticality, we propose for future study, the following analogy between sandpiles and log gravity. The relaxation of the Brown-Henneaux boundary conditions leading to the appearance of the logarithmic primary mode $\psi^{new}_{\mu \nu}$ at the critical point $\mu l =1$ corresponds to the dropping of the sand grain which leads to grain accumulation on a sandpile, bringing it to criticality by violating the local stability at a site, and the induced stochastic disturbance relaxes slowly in a multiscale cascade of events through a random distribution. Indeed, various equivalences between disordered systems and sandpile models have already been formulated \cite{wiese2022theory}. Furthermore, just as the log sector of log gravity was studied in terms of Feynman diagrams, the sandpile model was extensively studied in the case of embedded graphs \cite{volchenkov2009renormalization}. Our point d' appui is the Fefferman-Graham expansion for the metric in Gaussian coordinates used in \cite{Grumiller:2008qz} to encompass the new mode $\psi^{new}_{\mu \nu}$ and the evolution-based role of the metric term $\gamma_{ij}^(1)$ involved in the growth mechanism of the log sector, to better understand the diffusion geometry and self-criticality of log gravity. This work brings evidence that log gravity possesses some non-trivial geometric properties through the appearance of hidden self-similar symmetries observed in the multiscale organization of the log sector. The self-similarity underlying the structure of the log sector at different scales could be explored using some geometric renormalization schemes inspired by RG concepts, involving selfsimilar sandpile configurations. We hope that by employing tools from algebraic geometry \cite{perkinson2011primer,kalinin2018self,lang2019harmonic,kalinin2020pattern}, some progress will be made in the aforementioned direction.


\paragraph{Acknowledgements} We thank the anonymous referee for comments and suggestions to improve the presentation of the paper. This work is supported by the South African Research Chairs initiative of the Department of Science and Technology and the National Research Foundation, and by the National Institute for Theoretical and Computational Sciences, NRF Grant Number 65212.

\clearpage

\bibliographystyle{utphys}
\bibliography{sample}
\end{document}